\documentclass[twocolumn,aps,pra,epsfig,graphics,showpacs]{revtex4}
\usepackage{graphicx}
\usepackage{amsmath}
\usepackage{amsfonts}
\usepackage{amssymb}

\newcommand{\pr}[1]{|#1\rangle\langle#1|}
\newcommand{\mL}{\mathcal{L}}

\begin{document}

\title{Quantum limited velocity readout and quantum feedback cooling of a trapped ion via electromagnetically induced transparency}
\author{P. Rabl, V. Steixner and P. Zoller}
\affiliation{Institute for Theoretical Physics, University of Innsbruck, and\\
Institute for Quantum Optics and Quantum Information of the
Austrian Academy of Science, 6020 Innsbruck, Austria\\}

\begin{abstract}
 We discuss continuous observation of the momentum of a single atom
 by employing the high velocity sensitivity of the index of refraction in a driven
$\Lambda$-system based on electromagnetically induced transparency (EIT). In the ideal limit of unit collection efficiency this provides a quantum limited
measurement with minimal backaction on the atomic motion. A feedback loop, which drives the atom with a force proportional to measured signal, provides a cooling
mechanism for the atomic motion. We derive the master equation which describes the feedback cooling and show that in the Lamb-Dicke limit the steady state
energies are close to the ground state, limited only by the photon collection efficiency. Outside of the Lamb-Dicke regime the predicted temperatures are well
below the Doppler limit.
\end{abstract}


\maketitle

\section{Introduction}

Quantum feedback control employs the strategy of acting on a system based on measurement data obtained by continuous observation of the quantum system of
interest, thus achieving control of quantum dynamics and preparation of particular quantum states
\cite{Feedback_PRL,Feedback_PRA,StateEstimation,OptimalFeedback_PRL}. A prerequisite of developing quantum feedback control is the realization of quantum limited
measurements. Quantum optical systems and, more recently, mesoscopic systems have taken a leading role in achieving these requirements towards demonstration of
feedback control in the laboratory
\cite{Rempe_Feedback,Mabuchi_Cavity,Vitali_IonFeedback,Walls_IonFeedback,Orozco_FeedbackTheory,Mabuchi_SpinSqueezing,Vitali_Mirror,Mirror_Experiment,Schwab_Feedback}.

Motivated by the remarkable experimental progress with trapped ions, we will develop in the present paper a theory of quantum feedback cooling of a single ion,
based on a continuous readout of the atomic velocity via {\em dispersive} interactions with laser light. The idea is to devise an (in principle) quantum limited
measurement of the velocity by employing the strong detuning (and thus velocity) dependence of the index of refraction of a driven $\Lambda$-system near the
atomic dark state. These dark states are coherent superpositions of two atomic ground states which do not couple to the excited atomic state, which leads to
strong suppression of dissipative light scattering. This is the same feature which underlies recent studies of electromagnetically induced transparency, slow
light in atomic gases and quantum memory of light in atomic ensembles. The present setup of dispersive readout of the atomic velocity complements and is in
contrast to ongoing experiments of quantum feedback cooling of a single two-level ion in front of a mirror \cite{Bushev_Diss,Blatt_Feedback}, where the position
of the ion is continuously monitored by emission of light into the mirror mode, as analyzed theoretically in our recent publication~\cite{Viktor_Mirror}.

Our discussion of quantum feedback cooling of a single trapped ion in a strongly driven atomic $\Lambda$-system builds on, and connects various well-developed
topics in atomic physics and quantum optics, as well as continuous measurement and quantum feedback theory. Thus we find it worthwhile to present both a brief
review of the background material and physical key ideas underlying the present work, as well as the main results of the paper in Section~\ref{sec:Overview}. In
Section~\ref{sec:Model} we give the technical details of our model and derive the equations for the measured signal and the conditioned evolution of the atomic
motion for a weakly excited atom. A Wiseman-Milburn-type~\cite{Feedback_PRL} master equation for feedback cooling and the resulting temperatures will be
discussed in Section~\ref{sec:Feedback}. Finally, in Section~\ref{sec:EITCooling} we make a connection to EIT-laser cooling and describe combination of feedback
and laser cooling. The details of the adiabatic elimination of the internal atomic states are given in the Appendixes~\ref{app:AE_General}
and~\ref{app:AE_LambDicke}.

\section{Overview and Summary}\label{sec:Overview}

In this section we present an overview of the concepts and the main results of this paper. Our emphasis will be on explaining the basic physics and strategy
behind our quantum feedback scheme, and providing references to later sections where the mathematically inclined reader can find the details of the derivations.
In Subsection \ref{subsec:EIT} we will briefly review EIT and discuss the dependence of the index of refraction on the ion momentum near the dark state
resonance. Continuous read out of the ion momentum using homodyne detection will be formulated in Subsection \ref{subsec:ContinuousMeasurement}. The main results
of the present paper are the equations for feedback cooling in Subsection \ref{part:Feedback}, and the predictions for the final temperatures and cooling rates
in \ref{subsec:QFCLD}, in particular also in connection to EIT laser cooling of ions~\cite{EITCooling_PRL,EITCooling_PRA,EITCooling_Exp}.

\subsection{Electromagnetically Induced Transparency}\label{subsec:EIT}

\begin{figure}
\begin{center}
\includegraphics[width=0.45\textwidth]{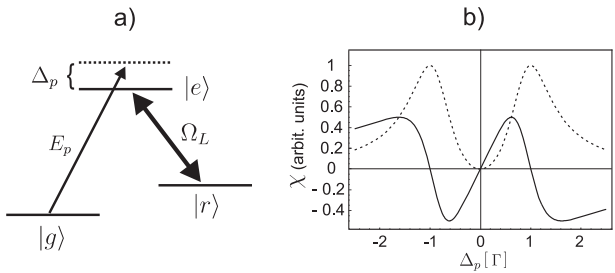}
\caption{a) Atomic $\Lambda$-system with two ground states
$|g\rangle$ and $|r\rangle$. The transition $|r\rangle \rightarrow
|e\rangle$ is driven by a strong resonant light field with Rabi
frequency $\Omega_L$, while the probe field $E_p$ couples
$|g\rangle \rightarrow |e\rangle$ with detuning $\Delta_p$. b)
Susceptibility $\chi(\Delta_p)$ of a $\Lambda$-system as a
function of the detuning of the probe field, $E_p$. At the center
of the transparency window the real part of the susceptibility
(solid line) representing the index of refraction of the medium
exhibits a steep slope, while the imaginary part (dotted line)
vanishes.} \label{fig:EIT_classic}
\end{center}
\end{figure}

The phenomenon of EIT is related to a quantum interference effect
which in its simplest form can be observed in a three level atom
(for a review see:~\cite{Scully_Book,Lukin_RMP} and references
therein). To discuss this effect we consider an atom with the
internal states $|g\rangle$, $|e\rangle$ and $|r\rangle$ in a
$\Lambda$-configuration as shown in Fig.~\ref{fig:EIT_classic}.
The transition between the states $|r\rangle$ and $|e\rangle$ is
driven by a strong, resonant laser, while a second light field,
the probe field, couples the ground state to the excited state. We
denote by $\Delta_p=\omega_p- \omega_{eg}$ the detuning of the
probe field from the atomic resonance. The atomic Hamiltonian is
then given by
\begin{equation} \label{eq:Hamil}
\begin{split}
H_\Lambda=\hbar \Delta_p \pr{g}+\frac{\hbar}{2}\left(\Omega_L
|e\rangle\langle r| + g |e\rangle\langle g|  + h.c. \right) \,,
\end{split}
\end{equation}
where $\Omega_L$ and $g$ are the Rabi frequencies of the laser and
the probe field respectively.  At the two photon resonance,
$\Delta_p=0$, the Hamiltonian, $H_\Lambda$ has an adiabatic
eigenstate with zero energy, a so-called ``dark state'',
\begin{equation}
|D\rangle\sim \Omega_L|g\rangle -g |r\rangle\,.
\end{equation}
For an atom in the state $|D\rangle$, the excitation from the
states $|g\rangle$ and $|r\rangle$ destructively interfere and the
atom decouples from the light. We note that an atom in a dark
state involves no excited state population, and is thus immune to
decay from the excited state.

The existence of such a dark state leads to remarkable properties
of the index of refraction. For weak probe fields, the propagation
can be discussed in terms of the linear susceptibility,
$\chi(\omega_p)$. The real part of $\chi(\omega_p)$ is related to
the refractive index by $n=1+{\rm Re}(\chi(\omega_p))/2$, while
the imaginary part, ${\rm Im}(\chi(\omega_p))$, is proportional to
the absorption coefficient. For an ensemble of three level atoms
where both, $|g\rangle$ and $|r\rangle$ are long-lived states, the
susceptibility has the characteristic behavior~\cite{Lukin_RMP},
\begin{equation} \label{eq:chi}
\chi(\Delta_p)\sim\frac{i\Delta_p}{|\Omega_L|^2/4+i\Delta_p(\Gamma+i\Delta_p)
}\,,
\end{equation}
where $\Gamma$ is the decay rate of the excited state.
Fig.~\ref{fig:EIT_classic}b shows the dependence of $\chi$ on the
detuning, $\Delta_p$. Around the dark resonance $\Delta_p=0$ we
have a steep slope of the refractive index (solid line) which
leads to a slow group velocity of the probe field (``slow light'')
while absorption is strongly suppressed (dotted line), giving rise
to ``electromagnetically induced transparency''. The width of the
transparency window as well as the variation of the refractive
index depend on $\Omega_L$ and can be controlled by the laser
field. We note the suppression of absorption at the dark state
resonance (dotted line in Fig.~\ref{fig:EIT_classic}b).

\begin{figure}
\begin{center}
\includegraphics[width=0.40\textwidth]{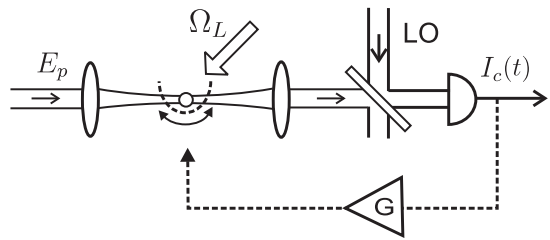}
\caption{Schematic setup for the continuous observation of the atomic momentum. The trapped atom is driven by a strong laser to create the transparency effect
for the probe field, $E_p$. Modulations of the probe light are detected in a homodyne measurement, i.e., by mixing $E_p$ with the strong field of the local
oscillator (LO). For feedback cooling, a force proportional to the measured signal, $I_c(t)$, is applied on the atom. Alternatively to the lens system, the atom
can be placed inside a running wave cavity, e.g., to enhance the photon collection efficiency, $\epsilon$.} \label{fig:EITSetup}
\end{center}
\end{figure}

Consider now a  single ion in a $\Lambda$-configuration
(Fig.~\ref{fig:EITSetup}a)
 moving in a trapping potential. We will only consider the 1D motion along the
propagation direction of both the probe and dressing laser beams.
If we adopt for the moment a classical description of the ion
motion with $z(t)$ the ion trajectory, then the internal dynamics
of the ion can again be described by the Hamiltonian
(\ref{eq:Hamil}) with a Doppler-shifted probe detuning
\[
\Delta_p(t)=\Delta_p + (k_p -k_L)\cdot v(t)\,,
\]
where  $k_p$ and $k_L$ the wave vectors of the running probe field
and the dressing laser field, respectively, and $ v(t) \equiv \dot
z (t)$  is the ion velocity. Thus for a resonant probe field,
$\Delta_p=0$, the change in the index of refraction is a linear
function of the atomic velocity or momentum with the steep slope
given in Eq.~\eqref{eq:chi}.

This suggests the strategy to measure the momentum of the ion
continuously by monitoring the phase shift due to the varying
index of refraction. We note that - while the index of refraction
of a single particle is small - EIT will strongly amplify the
sensitivity to the velocity. At the same time dissipation due to
light scattering is strongly suppressed within the transparency
window. These arguments can be easily adapted to a situation where
the atomic motion is quantized.

\subsection{Continuous observation of the momentum of the trapped ion}
\label{subsec:ContinuousMeasurement}

We turn now to a formulation of the continuous read out of the ion
momentum as outlined in Fig.~\ref{fig:EITSetup}. The idea is to
measure the momentum of the ion via the phase changes of the probe
beam as described in the previous subsection. The phase of the
probe beam can be determined by homodyning, i.e.~by mixing the
probe beam with a local oscillator and measuring the homodyne
current, $I_c(t)$.

The state of the observed system (the moving ion) is described by
a conditional density operator $\mu_c(t)$, which represents the
observer's knowledge of the current state of the system for a
given record of the measured signal, $I_c(t)$. We will show in
Sec. \ref{sec:Model} that after adiabatic elimination of the
excited state of the (weakly driven) ion the measured homodyne
current has the form,
\begin{equation}\label{eq:Current}
I_c(t)=2\epsilon\Gamma_0 \langle\hat p\rangle_c +
\sqrt{\epsilon\Gamma_0}\,\xi(t)\,.
\end{equation}
The current is the sum of two terms.  The first contribution shows a linear dependence on the conditional expectation value of the momentum operator $\langle\hat
p\rangle_c \equiv {\rm Tr}\{\mu_c \hat p \}$. Thus by measuring $I_c(t)$ we learn the momentum of the moving ion.  The second contribution describes a shot noise
term with $\xi(t)$ a white noise Gaussian process. The signal strength is determined by the rate $\Gamma_0$. It is related to the slope of the refractive index
$\chi$ at $\Delta_p=0$: an explicit expression is given in Sec.~\ref{sec:Model} Eq.~\eqref{eq:PrelimCurrent} below. The parameter $0<\epsilon\le 1$ takes into
account the collection efficiency of the scattered photons (where in the ideal case $\epsilon = 1$). Eq.~\eqref{eq:Current} is derived under  the assumption that
$\Omega_L$ is large compared to typical Doppler detuning, $\Delta_D$, and is valid on time scale which is slow compared to $\Omega_L^{-1}$. The signal is
maximized for the local oscillator phase, $\phi=0$.

According to continuous measurement theory applied to homodyne
detection, the conditional density operator $\mu_c(t)$ is updated
upon observation of the current $I_c(t)$ following the Ito
equation,
\begin{eqnarray}\label{eq:ME}
d\mu_c(t)&=&-i\nu [\hat a^\dag \hat a,\mu_c(t)]dt \\
 &&+\mathcal{L}_M
\mu_c (t)\,dt+\sqrt{\epsilon\Gamma_0}\mathcal{H}[\hat p]\mu_c
(t)\,dW(t)\,. \nonumber
\end{eqnarray}
This equation will be derived in Sec.~\ref{sec:Model}, Eq.~\eqref{eq:PrelimSME}.

The first term in this equation describes the free evolution in the 1D harmonic trap where $\nu$ denotes the trap frequency, and $\hat a$ ($\hat a^\dagger$) the
destruction (creation) operators, respectively. The effects of the continuous observation appear in the second and third term of Eq.~\eqref{eq:ME}.

The superoperator $\mL_M$ determines the back action of the measurement setup on the atomic motion. In the Lamb-Dicke limit $\eta=2 \pi a_0/\lambda_p \ll 1$,
where the extension of the atomic wavepacket (size of the harmonic oscillator ground state $a_0$) is much smaller than the wavelength of the light, $\lambda_p$,
it has the form,
\begin{equation}\label{eq:L}
\mathcal{L}_M\mu=-\frac{\Gamma_0}{2}[\hat p,[\hat p,\mu]]\,.
\end{equation}
The action of $\mL_M$ tends to diagonalize the density operator in
the eigenbasis of the (measured) operator $\hat p$.  By comparing
the decoherence rate, $\Gamma_0$, with the signal strength,
$\epsilon \Gamma_0$, we see that for $\epsilon<1$ the measurement
is not quantum limited, i.e., more noise is added than required by
quantum mechanics~\cite{Braginsky}. Although the measurement does
not reach the quantum limit, the back action is still minimal for
a given collection efficiency.

In the third term of Eq.~\eqref{eq:ME} we introduced the notation,
\[
\mathcal{H}[\hat c]\mu=\hat c \mu + \mu \hat c -\langle c+\hat c
\rangle \mu \,.
\]
This term describes the observer's knowledge of the current state
of the system and therefore depends on the measured signal. The
Wiener increment $dW(t)$ is formally related to the signal noise
by $dW(t)\equiv \xi(t)dt$.

In summary, Eq.~(\ref{eq:Current}) for the homodyne current
$I_c(t)$ and the evolution equation (\ref{eq:ME}) for the
conditional density matrix constitute the basic equations of
continuous observation of the momentum of the ion via homodyne
detection.

\subsection{Quantum Feedback Cooling}\label{part:Feedback}

The information on the atomic momentum contained in the signal $I_c(t)$ (\ref{eq:Current}) can be used to act back on the system. Here we are interested in
cooling the atomic motion by using the feedback strategy known as ``cold damping''~\cite{MirrorQuiescence}. The idea is to apply a force on the atom which is
proportional but opposite to its momentum. This force creates an effective friction for the atomic motion and, therefore, leads to a dissipation of kinetic
energy.

In our setup the measured signal is already proportional to the
average momentum, $\langle \hat p \rangle_c$, and can be amplified
and fed back directly. Thus we consider a feedback Hamiltonian of
the form,
\begin{equation}\label{eq:Hfb}
H_{fb}(t) =  \frac{G}{2\epsilon} I_c(t-\tau) \hat z\,,
\end{equation}
where $G$ is the dimensionless gain factor, and $\tau$ is the
finite delay in the feedback loop.  Note that $\tau>0$, so that
$H_{fb}$ acts after the measurement. Eq.~(\ref{eq:ME}) with the
feedback Hamiltonian added provides us with a {\em feedback
equation} describing the time evolution of the system. The goal is
now to average this equation over the Gaussian white noise
$\xi(t)$.

A general theory for direct quantum feedback has been first discussed in a seminal paper by Wiseman and Milburn~\cite{Feedback_PRL}. In particular, they have
shown how to average the quantum feedback equation in the limit $\tau \rightarrow 0^+$. In our case this assumption implies that the time delay of the feedback
is small on the scale of the (adiabatically eliminated) system evolution, a condition which is realistic in the present context. Adopting this formalism we will
derive in Sec.~\ref{sec:Feedback} a master equation for the unconditioned density operator, $\mu(t)=E[\mu_c(t)]$. We obtain (for $\Delta_p=\Delta_L=0$)
\begin{equation}\label{eq:FeedbackME}
\begin{split}
\dot \mu=&-i\nu [ \hat a^\dag \hat a,\mu] +\mathcal{L}_M\mu\\
&-i\Gamma_0\frac{G}{2}[\hat z, \hat p \mu+\mu\hat
p]-\Gamma_0\frac{G^2}{8\epsilon}[\hat z,[\hat z,\mu]]\,.
\end{split}
\end{equation}
The first line of this equation describes the free evolution and the measurement back action, $\mathcal{L}_M$ Eq.~(\ref{eq:L}). The terms in the second line of
Eq.~\eqref{eq:FeedbackME} include the effects of the feedback loop. While the term proportional to $G$ causes the expected damping of the motion, the second term
leads to a diffusion of the atomic momentum. This diffusion originates from the noise in the measured current which is also amplified and fed back to the system.

\begin{figure}
\begin{center}
\includegraphics[width=0.35\textwidth]{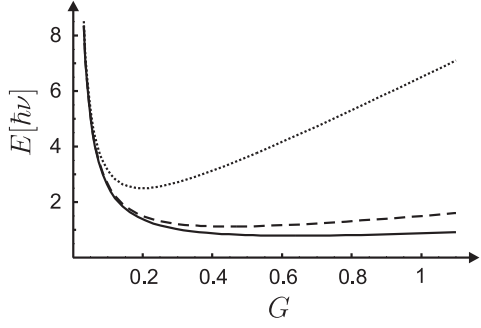}
\caption{The figure shows the dependence of the steady state
energy as a function of the feedback gain, $G$. The results are
calculated in the Lamb-Dicke limit and for $\Gamma_0=0.01\,\nu$.
The three curves are plotted for the parameters, $\epsilon=0.1$
(solid line), $\epsilon=0.05$ (dashed line) and $\epsilon=0.01$
(dotted line). } \label{fig:GainPlot}
\end{center}
\end{figure}

The generic dependence of the steady state energy on the feed back
gain calculated from Eq.~\eqref{eq:FeedbackME} is plotted in
Fig.~\ref{fig:GainPlot}. The curves show the expected minimum as a
function of $G$, at the point where the noise of the feedback loop
starts to dominate over the damping force. More detailed results
will be presented in the following subsection.

\subsection{Results: Quantum Feedback vs. EIT Laser Cooling}\label{subsec:QFCLD}

In general, the interaction of atoms with light always leads to some form of laser cooling or heating. In the resonant case $\Delta_p=\Delta_L=0$ (discussed
above), which is required to measure the atomic momentum, heating and cooling rates are equal and cause the diffusion described by $\mathcal{L}_M$.

\begin{figure}
\begin{center}
\includegraphics[width=0.35\textwidth]{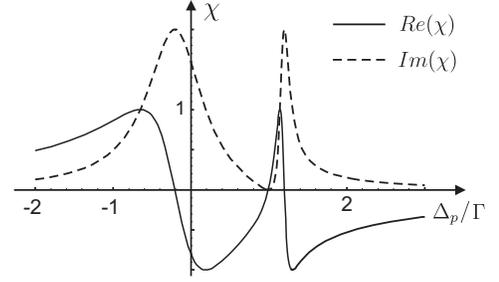}
\caption{Linear susceptibility, $\chi(\Delta_p)$, in arbitrary
units for $\Delta_L/\Gamma=\Omega_L/\Gamma=1$.  }
\label{fig:EITSusceptibility}
\end{center}
\end{figure}

By detuning the lasers away from the resonance, $\Delta_L\neq0$,
the atomic susceptibility, $\chi(\Delta_p)$, and therefore the
absorption properties become quite asymmetric (see
Fig.~\ref{fig:EITSusceptibility}). This asymmetry is exploited in
EIT laser cooling (ELC), where the absorption on the red sideband
(a phonon is removed from the motion) is much more likely than on
the blue sideband (a phonon is added to the motion). A detailed
discussion of ELC can be found in Ref.~\cite{EITCooling_PRA}.

In Sec.~\ref{sec:EITCooling} we show that in the Lamb-Dicke limit we can derive the extension of Eq.~\eqref{eq:FeedbackME} which includes quantum feedback
cooling as well as ELC. In rotating frame with respect to the trap frequency, $\nu$, it can be written in the form,
\begin{equation}\label{eq:FeedbackME_LD}
\begin{split}
\dot \mu = (A_-+A_-^{fb})\mathcal{D}[\hat a]\mu+
(A_++A_+^{fb})\mathcal{D}[\hat a^\dag]\mu\,,
\end{split}
\end{equation}
with
\[
\mathcal{D}[\hat a]\mu=\hat a \mu \hat a^\dag -\frac{1}{2}\hat
a^\dag \hat a \mu -\frac{1}{2}\mu \hat a^\dag \hat a \,.
\]

The total cooling and the total heating rate are divided into a contribution from the atom-laser interaction, $A_\pm$, and a contribution from the feedback
force, $A_\pm^{fb}$. They are given by,
\begin{eqnarray*}
A_\pm&=&\frac{\Gamma_0}{2}{\rm Re}[I(\pm \nu)]\,,\\
 A_\pm^{fb}&=&\Gamma_0\left(G\frac{\nu \Gamma}{\Omega^2}{\rm Im}[I^*(\pm \nu)e^{i\phi}]+\frac{G^2}{8\epsilon}\right)\,.
\end{eqnarray*}
The relation between the four rates is determined by the function $I(\nu)$ which is defined as
\begin{equation}\label{eq:I}
I(\nu)=\frac{\Omega^4}{2\Gamma\nu^2}\frac{i\nu}{(\Omega^2-4\nu(\nu-\Delta_L))+i2\Gamma\nu}\,.
\end{equation}
Note that the rates $A_\pm$ are proportional to the imaginary part of the atomic susceptibility at the sideband frequencies, $\chi(\omega_p\mp\nu)$, (see
Fig.~\ref{fig:EITSusceptibility}). For a general detuning $\Delta_L$, the measured signal is no longer proportional to $\langle \hat p \rangle$, and therefore
the phase of the local oscillator, $\phi$, appears in the feedback rates.

In the basis of harmonic oscillator states the master
equation~\eqref{eq:FeedbackME_LD} has the form of a standard rate
equation for the trap occupations ($p_n=\langle n | \rho
|n\rangle)$ familiar from laser cooling,
\begin{equation}
\begin{split}
\dot p_n =&(A_-+A_-^{fb})\left[(n+1)p_{n+1}-n p_n\right] \\
          &+(A_++A_+^{fb})\left[(n-1)p_{n-1}-(n+1) p_n\right]\,,
\end{split}
\end{equation}
which predicts in steady state a Bose-Einstein distribution with a mean occupation number
\begin{equation}
\bar n = \frac{A_++A_+^{fb}}{(A_--A_+)+(A_-^{fb}- A_+^{fb})}\,.
\end{equation}

We now turn to the discussion of results for the case of pure
feedback cooling, and combined feedback and laser cooling:

\emph{Pure feedback cooling}. We first reproduce the results of Eq.~\eqref{eq:FeedbackME} by setting $\Delta_p=\Delta_L=0$. Then $A_\pm=\Gamma_0/2$ and the
cooling of the atom is attributed to the feedback mechanism. The minimal energy is reached for $G=\sqrt{4\epsilon}$ and $\phi=0$. It is given by
\begin{equation}\label{eq:Emin}
E_{\rm min}=\frac{\hbar \nu}{2}\sqrt{\frac{1}{\epsilon}}\,.
\end{equation}
This expression shows that the final temperature is only limited
by the collection efficiency $\epsilon$. In the theoretical limit,
$\epsilon\rightarrow 1$, it approaches the ground state energy,
$\hbar\nu/2$.

\begin{figure}
\begin{center}
\includegraphics[width=0.35\textwidth]{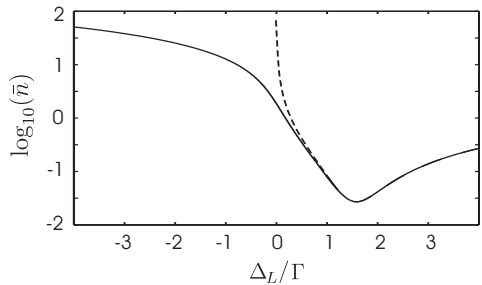}
\caption{Steady state occupation number, $\bar n=\langle \hat a^\dag \hat a\rangle$, as a function of the detuning, $\Delta_L$. The solid line shows the result
for optimized $\phi$ and optimized gain, $G$, while the dashed line shows the result without the feedback loop. The curves are plotted for the parameters:
$\epsilon=0.05$, $\Omega_L/\Gamma=0.8$ and $\nu/\Gamma=0.1$. } \label{fig:LambDicke}
\end{center}
\end{figure}

\emph{Feedback cooling and ELC}. When we tune away from the resonance, $\Delta_p=\Delta_L\neq 0$, we obtain a difference in the laser cooling rates, $A_-\neq
A_+$. In Fig.~\ref{fig:LambDicke} we compare the final temperatures for the optimal feedback gain with the case of pure ELC. For blue detuning, $\Delta_L>0$ the
mechanism of ELC sets in and cools the atom close to the ground state. Although the addition of the feedback loop always leads to even lower temperatures, its
effect can be neglected because for the present parameters ELC already provides efficient ground state cooling. For red detuning $\Delta_L\leq 0$ the absorption
spectrum (see Fig.~\ref{fig:EITSusceptibility}) is reversed and the atom is actively heated by the light absorption. In this case a steady state is only reached
when the feedback cooling dominates over the laser induced heating.

\section{The model}\label{sec:Model}

In this section we present a detailed description of our model for
the measurement setup which is shown in Fig.~\ref{fig:EITSetup}.
The system of interest is the three level atom which is confined
by an external trapping potential. This atom is illuminated by a
strong laser field to create the transparency effect for the probe
field. The outgoing probe light is mixed with a strong local
oscillator to perform a homodyne measurement to detect linear
shifts of the field.

To describe the dynamics of the atom as well as the detection of
scattered field, we start with the total Hamiltonian,
\begin{equation}
H=H_{A}+H_{A-EM}+H_{EM}\,.
\end{equation}
It is the sum of the Hamiltonian for the external and internal
states of the atom, $H_A$, the free Hamiltonian for the
electromagnetic environment, $H_{EM}$, and the coupling between
the atom and the electromagnetic field, $H_{A-E}$.

For the internal level structure we consider a $\Lambda$-configuration as shown in Fig.~\ref{fig:EIT_classic}a. A classical laser field with frequency $\omega_L$
drives the transition between the excited state, $|e\rangle$ and the second ground or metastable state, $|r\rangle$. The Rabi frequency for this coupling is
denoted by $\Omega_L$. For the external dynamics of the atom we restrict ourselves to a one dimensional model, i.e., we assume that the atom is strongly confined
in the x and y directions. Along the z-axis, which coincides with the propagation direction of the probe beam, the atom is trapped by the external potential,
$V(z)$. Although it is not essential for the following discussion, we further assume that $V(z)$ is harmonic, with a trap frequency, $\nu$. This assumption
allows us to introduce the dimensionless position and momentum operators, $\hat z:=(\hat a+\hat a^\dag)/\sqrt{2}$ and $\hat p:=i(\hat a^\dag -\hat a)/\sqrt{2}$,
where $\hat a$ and $\hat a^\dag$ denote the usual annihilation and creation operators. With the definition of the external Hamiltonian, $H_E=\hbar \nu \hat
a^\dag \hat a$, and the notation $\sigma_{ij}=|i\rangle\langle j|$ the atom Hamiltonian is then given by
\begin{equation}
\begin{split}
H_A=H_E&+\hbar \omega_e \pr{e}+\hbar \omega_r\pr{r}\\
&+\frac{\hbar \Omega_L}{2}\left( e^{-i\omega_r t}e^{i\eta_L \hat z}\sigma_{er}+ h.c\right)\,.
\end{split}
\end{equation}
Here $\omega_{e,r}$ denote the eigenfrequencies of the corresponding states $|e\rangle$ and $|r\rangle$. The Lamb-Dicke parameter is defined as,
$\eta_r=k_L\sqrt{\hbar/m\nu}$, where $m$ is the mass of the atom and $k_L$ is the wave vector (projected on the z-axis) of the laser field.

The electromagnetic environment consist of a three dimensional set
of plane wave modes, labelled by their wave vector, $\vec k$ and
their polarization, $\lambda$. In terms of the corresponding
annihilation and creation operators, $\hat b_\lambda(\vec k)$ and
$\hat b^\dag_\lambda(\vec k)$, the free evolution is determined by
the Hamiltonian,
\begin{equation}
H_{EM}=\sum_{\lambda=1,2}\int d^3k\, \hbar \omega_{\vec k}\,  \hat
b^\dag_\lambda(\vec k) \hat b_\lambda(\vec k)\,.
\end{equation}
The electric field of the environment interacts  with the internal
states of the atom via a dipole coupling. Under the rotating wave
approximation the interaction Hamiltonian is
\begin{equation}
\begin{split}
H_{\rm A-E}=- \vec \mu_{eg} \vec E^+(\hat z)\sigma_{eg}-\vec
\mu_{er} \vec E^+(\hat z)\sigma_{er} + h.c. \,,
\end{split}
\end{equation}
with the standard expression for the electric field,
\begin{equation}
\vec E^+(\vec x)=i\sum_{\lambda=1,2}\int d^3 k \,
\sqrt{\frac{\hbar \omega_k}{2\epsilon_0(2\pi)^3}}\,\vec
\varepsilon_\lambda(\vec k) \hat b_\lambda(\vec k)\, e^{i\vec k
\vec x}\,.
\end{equation}
In an experiment the lens system defines a certain spatial mode function for the probe beam. To describe the homodyne detection of the probe field, we  divide
the total electric field into the field of this particular mode, $\vec E_p$, and a remaining set of modes, orthogonal to $\vec E_p$,
 \begin{equation}
\vec E^+(\vec x)= \vec E_p^+(\vec x)+ \vec E^+_\perp(\vec x)\,.
 \end{equation}
 In our model we approximate $\vec E_p$ by the one dimensional field,
\begin{equation}\label{eq:ProbeField}
\vec E_p^+(z)=i \vec{\mathcal{E}_p}\,e^{i(k_p z-\omega_p t)}
+i\vec \varepsilon_p\int_0^{\infty} dk \,a(k) \,e^{i k z} \hat
b_p(k)\,.
\end{equation}
The first part in this expression describes the coherent field of the incoming probe beam. Note that with this definition of the operator, $\vec E_p$, the
initial state of the electromagnetic environment is the vacuum state. The function, $a(k)$, determines the coupling of the atom to the dense set of modes, $\hat
b_p (k)$. It must be adjusted to reproduce the correct results of the real mode function.

The interaction of the atom with the coherent part of the probe field leads to transitions between $|g\rangle$ and $|e\rangle$, characterized by the Rabi
frequency, $g=2|\vec \mu_{eg}\vec{\mathcal{E}_p}|/\hbar$. In the following include this term in the atomic evolution and define a new system Hamiltonian,
$H_S=H_A+H_{A-EM}|_{\rm coh}$. In a frame rotating with the laser frequencies, $\omega_p$ and $\omega_L$, this Hamiltonian is then given by
\begin{equation}\label{eq:Hamiltonian}
\begin{split}
H_S=H_E &-\hbar \Delta_p\pr{e}-\hbar(\Delta_p-\Delta_L)\pr{r} \\
&+\frac{\hbar \Omega_L}{2}\left(e^{i\eta _r\hat z}\sigma_{er}+e^{-i\eta_r\hat z}\sigma_{re}\right)\\
&+\frac{\hbar g}{2}\left(e^{i\eta_g\hat
z}\sigma_{eg}+e^{-i\eta_g\hat z}\sigma_{ge}\right)\,.
\end{split}
\end{equation}
Here we introduced the detunings $\Delta_p=\omega_p - \omega_e$ and $\Delta_L=\omega_L-(\omega_e-\omega_r)$, and the Lamb-Dicke parameter of the probe field,
$\eta_g=k_p\sqrt{\hbar/m\nu}$.

The coherent evolution of the driven atom is now determined by the system Hamiltonian $H_S$.  The coupling between the atom and the dense set of modes of
 $\vec E_\perp$ and the non-classical part of $\vec E_p$ has two effects. First, it leads to an incoherent dynamic of the atomic state.
 This includes the decay of the excited state population as well as a diffusion of the atomic momentum due to the random recoil kicks of the emitted
photons.  Second, the coupling changes the state of the
electromagnetic environment which is (partially) detectable by the
observer.

\subsection{Master equation}
 We first look at the incoherent dynamics of the atom and ignore the evolution of the electromagnetic bath. By applying the standard Born-Markov
approximation and tracing over the bath degrees of freedom we obtain a master equation for the system density matrix, $\rho$. It can be written in the standard
form, $\dot \rho=\mathcal{L}\rho$, with a Liouville operator~\cite{Stenholm_RMP,PZ_IonLaserCooling,EITCooling_PRA},
\begin{equation}\label{eq:GeneralME}
\mathcal{L}\rho=-\frac{i}{\hbar}[H_{\rm eff}\rho- \rho H_{\rm
eff}^\dag]+\mathcal{J}_g(\sigma_{ge}\rho\sigma_{eg})+\mathcal{J}_r(\sigma_{re}\rho\sigma_{er})\,.
\end{equation}
The effective Hamiltonian, $H_{\rm eff}=H_S-i\hbar\Gamma/2 \pr{e}$, includes the unitary evolution of the atom as well as the decay of the excited state
population with a total rate, $\Gamma$. The two ``recycling'' terms are defined by
\begin{equation}
\mathcal{J}_{j=r,g}(\rho)=\Gamma_j \int_{-1}^1 du N(u)
\,e^{-i\eta_j\hat z u} \rho  e^{i\eta_j\hat z u}\,.
\end{equation}
The rates $\Gamma_g$ and $\Gamma_r$  denote the decay rates into the corresponding states, $|g\rangle$ and $|r\rangle$. The dipole distribution,
$N(u)=\frac{3}{8}(1+u^2)$, with $u=\cos(\varphi)$, determines the probability for emitting a photon under a certain angle, $\varphi$, with respect to the z-axis.

 Master equation~\eqref{eq:GeneralME} describes the full dynamics of a driven atom in a $\Lambda$-configuration. The external and internal degrees of freedom are
 coupled via the position dependent interaction with the electromagnetic field. In general, the recoil kicks of the emitted photons described by $\mathcal{J}_{g,r}$
 lead to momentum diffusion and a heating of the atomic motion. For an appropriate choice of laser detunings the heating can be compensated by photon absorptions
 (laser cooling). In this paper, we follow a different approach where cooling is provided by an external feedback force.

\subsection{Continuous homodyne detection}
In a next step we describe the homodyne detection of the probe
field, $\vec E_p$.
 Here we follow the standard theory of homodyne detection (see, e.g. Ref.~\cite{QuantumNoise,Carmicheal_OpenSystems,Wiseman_Homodyne}) to derive the
 relevant equations for the model specified above.


 After the interaction with the atom, the outgoing probe field is desribed by the Heisenberg operator $\hat b_{p,{\rm out}}(t)$.
 Using the input-output formalism~\cite{QuantumNoise} it is related to the incoming field, $\hat b_{p,in}(t)$, by
\begin{equation}\label{eq:Outfield}
 \hat b_{p,{\rm out}}(t)=\hat b_{p,in}(t)+\sqrt{\gamma}\hat c_p(t)\,,
\end{equation}
where $\hat c_p=e^{-i\eta_p \hat z}\sigma_{ge}$ denotes the atomic ``jump operator'' which couples to the probe field. To relate our model to the real
experimental setup, we set $\gamma= \epsilon \Gamma_g$, where the collection efficiency $\epsilon$ determines the fraction of photons which are scattered into
the mode of the probe field.

In homodyne detection the outgoing field~\eqref{eq:Outfield} is
mixed with a strong coherent field of the local oscillator. When
the transmittance of the beam splitter is close to one the field
operator at the position of the detector is
\begin{equation}
\hat b_d(t)=  \sqrt{\gamma}\beta e^{i\phi}e^{-i\omega_p t}+\hat
b_{p,in}(t)+\sqrt{\gamma}\hat c_p(t)\,.
\end{equation}
Here, $\beta$ and $\phi$ denote the real amplitude and the phase of the reflected part of the local oscillator. Note that the total probe field defined in
Eq.~\eqref{eq:ProbeField} is the sum of a classical and a quantized contribution. The classical part of $\hat b_{p,in}(t)$ can simply be absorbed in a
redefinition of $\beta e^{i\phi}$. The operator for the homodyne current is,
\begin{equation}
\hat I_h(t)=\lim_{\beta\rightarrow\infty}\left(\hat b_d^\dag(t)
\hat b_d(t) - \gamma \beta^2\right)/\beta\,.
\end{equation}
The measured signal $I_c(t)$ is then defined as the outcome of the continuous measurement of the current operator, $\hat I_h(t)$. Using the results form the
theory of homodyne detection~\cite{QuantumNoise}, we obtain
\begin{equation}\label{eq:ExactCurrent}
I_c(t)=\epsilon \Gamma_g\langle\hat c_p e^{-i\phi}+\hat c_p^\dag
e^{i\phi}\rangle_c(t) +\sqrt{\epsilon \Gamma_g}\xi(t)\,.
\end{equation}
The unconditioned evolution given by master equation~\eqref{eq:GeneralME} and the measurement record, $I_c(t)$, determine the evolution of the conditioned
density operator, $\rho_c(t)$. Following Ref.~\cite{Wiseman_Homodyne} we obtain,
\begin{equation}\label{eq:ExactME}
d\rho_c(t)=\mathcal{L}\rho_c(t)dt+\sqrt{\epsilon\Gamma_g}\mathcal{H}[\hat
c_p e^{-i\phi}]\rho_c(t)\,dW(t) \,.
\end{equation}

Eqs.~\eqref{eq:ExactCurrent} and~\eqref{eq:ExactME} represent a
full description of the conditioned dynamics of the atom under
continuous observation and serve as the starting point for the
following discussion.

\subsection{Adiabatic Elimination}\label{subsec:AE}
The current $I_c(t)$ as given in Eq.~\eqref{eq:ExactCurrent} is
still a function of the coupled external and internal states of
the atom. In the following we show that for a weakly excited atom
we can eliminate the dynamics of the internal states, and the
measured signal becomes a linear function of the atomic momentum
as given in Eq.~\eqref{eq:Current}. In addition we derive the
resulting back action of the measurement on the motional state of
the atom.

As already noted in Section~\ref{sec:Overview}, the phase shift of
the probe light is a linear function of the atomic momentum as
long as the typical Doppler detuning,
$\Delta_D=\nu\eta_p\langle\hat p\rangle$, is small compared to the
width of the transparency window, $\Omega_L$. Therefore, we can
apply perturbation theory in the parameter $\Delta_D/\Omega_L$ to
derive an effective equation for the external state. As a first
step in our calculation we perform a unitary transformation,
\begin{equation}\label{eq:UnitaryTransformation}
U=e^{i\eta_p \hat z\pr{e}}e^{-i\bar \eta \hat z\pr{r}}\,,
\end{equation}
where we set $\bar \eta:=\eta_p-\eta_L$. In the new basis the
system Hamiltonian  $\tilde H_S= U^\dag H_S U$ is given by
\begin{equation}
\begin{split}
\tilde H_S=H_E -&\hbar\left(\Delta_p-\nu\eta_p\hat p-\nu \eta_p^2/2\right)\pr{e}\\
-&\hbar\left( \Delta_p- \Delta_L+\nu\bar \eta \hat p-\nu \bar \eta^2/2\right)\pr{r} \\
+&\frac{\hbar \Omega_L}{2}\left(\sigma_{er}+\sigma_{re}\right)
+\frac{\hbar g}{2}\left(\sigma_{eg}+\sigma_{ge}\right)\,.
\end{split}
\end{equation}
 As in the classical case  (see Section~\ref{subsec:EIT}) the position dependence of the atom-laser
coupling is transformed into a frequency shift for the states
$|e\rangle$ and $|r\rangle$. In addition to the Doppler detunings,
$\nu \eta_p \hat p$ and $\nu \bar \eta \hat p$, the internal
states are also shifted by the appropriate recoil frequencies.
They account for the fact, that each absorption or emission of a
photon also transfers kinetic energy to the atom.

For the particular choice of laser detunings, $\Delta_p= \Delta_L+\nu\bar \eta^2/2$,  and in the absence of a trapping potential the Hamiltonian $\tilde H_S$ has
a dark eigenstate, $|\psi\rangle_D=|p=0,D\rangle$. Here $|p=0\rangle$ is the zero momentum eigenstate and $|D\rangle$ denotes the internal dark state,
\begin{equation}
|D\rangle=\left(g|r\rangle-\Omega_L|g\rangle\right)/\Omega\,,
\end{equation}
with $\Omega=\sqrt{\Omega_L^2+g^2}$. In the following we assume that this relation between the detunings is fulfilled. The system Hamiltonian can then be written
as
\begin{equation}
\tilde H_S=H_E+H_I+H_{\rm int}\,,
\end{equation}
such that $H_E$ and $H_I$ act on the external or the internal
states only, while $H_{\rm int}$ describes the coupling between
them,
\begin{equation}
H_{\rm int}=\hbar \nu \eta_p \hat p \pr{e}-\hbar \nu \bar \eta \hat p \pr{r}\,.
\end{equation}
With the definition $\Delta=\Delta_p-\nu \eta_p^2/2$ the
Hamiltonian of the internal states, $H_I$, reduces to the  one  of
a driven $\Lambda$-system at the two photon resonance,
\begin{equation}
H_I=-\hbar \Delta \pr{e}+\frac{\hbar
\Omega_L}{2}\left(\sigma_{er}+\sigma_{re}\right) +\frac{\hbar
g}{2}\left(\sigma_{eg}+\sigma_{ge}\right)\,.
\end{equation}

 In the limit where
the external and internal degrees of freedom decouple, $H_{\rm int}\rightarrow 0$, the conditioned dynamics determined by Eq.~\eqref{eq:ExactME} leads to a
relaxation of the atom into the state
\begin{equation}
\tilde \rho_c(t\rightarrow \infty)=\tilde \mu_0\otimes \pr{D}\,,
\end{equation}
where $\tilde \mu_0$ is an undetermined  state of the external degrees of freedom. The interaction with the atomic motion, $H_{\rm int}$, or to be more precise
the term $\hbar \nu \bar{\eta} \hat p\pr{r}$ couples the state $|D\rangle$ to the bright (internal) eigenstates of $H_I$, $|+\rangle$ and $|-\rangle$. They are
given by
\begin{eqnarray*}
|+\rangle&=&\cos(\theta)|e\rangle + \sin(\theta)\left(g|g\rangle +\Omega_L |r\rangle\right)/\Omega\,,\\
|-\rangle&=&\sin(\theta)|e\rangle - \cos(\theta)\left(g|g\rangle
+\Omega_L |r\rangle\right)/\Omega\,,
\end{eqnarray*}
where the mixing angle $\theta$ is defined by the relation
$\tan(\theta)=\Omega/(\sqrt{\Omega^2+\Delta^2}-\Delta )$. Theses
two states are separated from the dark state by the energies
\begin{equation}
\hbar \Omega_\pm=- \hbar(\Delta\mp \sqrt{\Omega^2+\Delta^2})/2\,.
\end{equation}
Therefore, for small Doppler shifts the population in the bright
states is of the order of $\Delta_D^2/\Omega_{\pm}^2$.

In the following we consider the limit where the eigen frequencies
of the bright states, $\Omega_\pm$, are much larger than the
typical Doppler detuning, $\Delta_D$, as well as the frequency of
the trap, $\nu$. The first assumption says that the atom is only
weakly excited and the total density operator is well approximated
by,
\begin{equation}
\tilde \rho_c(t)\simeq \tilde \mu_0(t)\otimes \pr{D}\,.
\end{equation}
The second condition, $\Omega_L\gg \nu$, ensures that the internal
state of the atom adiabatically follows the evolution of the
atomic momentum.
If both conditions are satisfied we can adiabatically eliminate the population in the the bright states and derive an effective equation for the evolution of the
motional state, $\tilde \mu_0(t)$. The details of this calculation are summarized in Appendix~\ref{app:AE_General}. Finally, we revert the unitary
transformation, $U$~\eqref{eq:UnitaryTransformation}, and trace over the internal states. For the resulting conditioned density operator of the motional state,
$\mu_c:= {\rm Tr}_I\{U\tilde \rho_c U^\dag\}$, we obtain the stochastic master equation,
\begin{equation}\label{eq:PrelimSME}
\begin{split}
d\mu_c(t)=&-i[\nu \hat a^\dag \hat a-\lambda^2\Delta \hat p^2,\mu_c(t)]dt +\mathcal{L}_M\mu_c(t) \,dt\\
&+\sqrt{\epsilon\lambda^2\Gamma_g}\mathcal{H}[\hat p e^{-i\phi}]\mu_c(t) dW(t)\,,
\end{split}
\end{equation}
and the homodyne current
\begin{equation}\label{eq:PrelimCurrent}
I_c(t)=2\epsilon\lambda^2\Gamma_g \cos(\phi)\langle\hat p\rangle_c + \sqrt{\epsilon\lambda^2\Gamma_g}\xi(t)\,.
\end{equation}
In these two equations we defined the parameter, $\lambda=2g\bar
\eta \nu/\Omega^2$,  and the average is take with respect to the
conditioned motional state, $\langle\cdot \rangle_c={\rm
Tr}\{\mu_c \cdot\}$. The measurement back action has the form
\begin{equation}\label{eq:Backaction}
\mathcal{L}_M
\mu=\frac{\Gamma_0}{2}\left[\frac{2\tilde{\mathcal{J}}_g}{1-\tilde{\mathcal{J}}_r}(\hat
p \mu \hat p) -\hat p^2 \mu -\mu \hat p^2\right]\,,
\end{equation}
with $\Gamma_0=\lambda^2\Gamma$ and
\begin{equation}\label{eq:RecyclingTerms}
\tilde{\mathcal{J}}_{j=g,r}(\mu)=\frac{\Gamma_j}{\Gamma}\int_{-1}^1
du N(u) e^{-i\eta_j \hat z (u-1)} \mu \, e^{i\eta_j \hat z
(u-1)}\,.
\end{equation}

\emph{Discussion}. The results given in the Eqs.~\eqref{eq:PrelimSME} and~\eqref{eq:PrelimCurrent} are valid in the limit of a weak probe field, $g\ll\Omega_L$.
In that case the dark state, $|D\rangle$, almost coincides with the ground state, $|g\rangle$, and the signal strength is maximized for a given strength of the
measurement back action, $\Gamma_0$, (see Appendix~\ref{app:AE_General}).  The signal, $I_c(t)$, can further be optimized setting $\phi=0$ and by choosing atomic
states with a small branching ratio $\Gamma_r/\Gamma_g$. The remaining difference between $\lambda^2 \Gamma_g$ and $\Gamma_0$ can be absorbed into the definition
of $\epsilon$. For $\Delta=0$ we then end up with Eq.~\eqref{eq:Current} and Eq.~\eqref{eq:ME} as given in Section~\ref{subsec:ContinuousMeasurement}.

\section{Feedback Cooling}\label{sec:Feedback}
As already mentioned in Section~\ref{sec:Overview} the goal of the
continuous momentum observation is to use the information in the
signal to manipulate the motion of the atom, e.g., to cool it. In
this section we discuss the implementation of the ``cold damping''
feedback strategy. By applying the theory of direct quantum
feedback~\cite{Feedback_PRL} we derive a master equation for the
unconditioned state, $\mu(t)=E[\mu_c(t)]$.

 For the feedback cooling we consider the measurement setup as described in the previous section. In addition, we apply a force on the atom which is proportional but
opposite to the measured current. For a single trapped ion, such a
feedback loop can be realized by converting the homodyne current
into a voltage difference between two trap electrodes. The effect
of the feedback loop on the system evolution can be written in the
general form,
\begin{equation}\label{eq:FeedbackContribution}
\dot \mu_c(t)|_{fb} = I_c(t-\tau)\mathcal{K}\mu_c(t)\,.
\end{equation}
The time delay of the feedback loop, $\tau$, can usually be neglected compared to the timescale of the atomic motion, $\nu^{-1}$. Nevertheless, for the
derivation of the final master equation a finite value of $\tau$ is important to obtain the correct operator ordering~\cite{Feedback_PRL}. To implement the idea
of ``cold damping'', we consider the feedback superoperator,
\begin{equation}
\mathcal{K}\mu=-i\frac{G}{2\epsilon}[\hat z, \mu]\,,
\end{equation}
where $G$ denotes the dimensionless gain factor. Note that with
this definition of $\mathcal{K}$, the frequency scale of the
feedback contribution is again of order $\Gamma_0$.

It has been shown in Ref.~\cite{Feedback_PRL} that
Eq.~\eqref{eq:FeedbackContribution} must be interpreted in the
Stratonovich sense. To be compatible with Eq.~\eqref{eq:ME} we
convert it into the Ito-type equation,
\begin{equation}\label{eq:ItoFeedbackConribution}
\begin{split}
d \mu_c(t)|_{fb}=&\Gamma_0\left(2\epsilon\langle \hat p \rangle_c(t-\tau)\mathcal{K}+\frac{\epsilon}{2}\mathcal{K}^2\right)\mu_c(t) dt\\
 &+\sqrt{\epsilon\Gamma_0}\,\mathcal{K}\mu_c(t)dW(t-\tau)\,.
\end{split}
\end{equation}
In this form we already see that the noise in the current $I_c(t)$ leads to the diffusion term $\mathcal{K}^2$. We can now add
Eq.~\eqref{eq:ItoFeedbackConribution} to the conditioned evolution given in Eq.~\eqref{eq:ME} and obtain the full stochastic dynamics of the atom under the
action of the feedback loop.

To derive a master equation which is independent of the measurement outcome, we perform an ensemble average over the stochastic process, $\xi(t)$ and obtain the
evolution of the unconditioned density operator, $\mu=E[\mu_c]$. When taking the average we must keep in mind that although $E[dW(t)]=0$, the Ito increment,
$dW(t-\tau)$, is not independent of $\mu_c(t)$, e.g., $E[dW(t-\tau)\mu_c(t)]\neq 0$. The way to perform the average in the limit, $\tau\rightarrow 0^{+}$, can be
found in Ref.~\cite{Feedback_Squeezing}. By following this procedure we end up with the master equation~\eqref{eq:FeedbackME} given in
Section~\ref{sec:Overview}.


\subsection{Feedback cooling in the Lamb-Dicke limit}
We first look at the solution of Eq.~\eqref{eq:FeedbackME} in the
Lamb-Dicke limit, $\eta_p,\eta_L \ll 1$. In this limit the recoil
kicks of the emitted photons can be neglected and the backaction
of the measurement~\eqref{eq:Backaction} simplifies to
\begin{equation}
\mathcal{L}_M\mu=-\frac{\Gamma_0}{2}[\hat p,[\hat p,\mu]]\,.
\end{equation}
For a harmonic trapping potential, $H_E=\hbar \nu \hat a^\dag \hat
a$, the feedback master equation~\eqref{eq:FeedbackME} is then
quadratic in the position and the momentum operators. Therefore,
the final state is Gaussian and we obtain analytic expressions for
the variances of $\hat z$ and $\hat p$. The resulting steady state
energy is given by
\begin{equation}
E=\frac{\hbar\nu}{2}\left(\frac{G\Gamma_0^2}{2\nu^2}+\frac{1}{G}+\frac{G}{4\epsilon}\right)\,.
\end{equation}
The first contribution  in the brackets originates from the
enhanced uncertainty of the position coordinate as a result of the
measurement of the momentum operator. If the measurement strength,
$\Gamma_0$, is much smaller than the trap frequency, $\nu$,
position and momentum coordinates are mixed sufficiently fast and
this contribution disappears. In this limit the optimal feedback
gain is given by $G=\sqrt{4\epsilon}$ and we obtain the minimal
energy given in Eq.~\eqref{eq:Emin}.

In Section~\ref{sec:EITCooling} we extend the discussion of the
feedback cooling in the Lamb-Dicke limit to arbitrary detunings,
$\Delta$. Then laser cooling effects play an important role for
the final temperatures.

\subsection{Feedback cooling beyond the Lamb-Dicke limit}

When the trapping potential is weak, the extension of the atomic
wavepacket can be of the order of the wavelength of the emitted
photons. In that case, the energy spacing in the trap, $\hbar
\nu$, is comparable to the recoil energy, $E_R$, and recoil kicks
from the emitted photons lead to an additional diffusion of the
atomic momentum. Therefore, we must take into account the full
expression for the back action term $\mathcal{L}_M$. By expanding
Eq.~\eqref{eq:Backaction} in the parameter $\Gamma_r/\Gamma$ we
can write it as
\begin{equation}
\mathcal{L}_M\mu=\frac{\Gamma_0}{2}\left(2\tilde{\mathcal{J}}_g \left(\sum_{n=0}^\infty \tilde{\mathcal{J}}_r^n \right)(\hat p \mu \hat p)- \hat p^2\mu - \mu
\hat p^2\right).
\end{equation}
The zeroth order term in the sum corresponds to the physical
picture where the atom is excited and simply decays back to the
ground state, $|g\rangle\approx|D\rangle$. Processes where the
atom first decays into the state $|r\rangle$, is then reexcited
again are taken into account by including higher order terms in
this sum.

In general, the full expression of $\mathcal{L}_M$ leads to a
hierarchy of coupled equations for the moments of $\hat p$ which
does not break off as in the Lamb-Dicke limit. In the following we
restrict our discussion to a finite trapping potential and
consider the limit, $\Gamma_0\ll\nu$.  As mentioned above, this
ensures a mixing of position and momentum coordinates and
therefore, an equal reduction of both variances. In this regime
non-energy conserving terms can be neglected and we obtain an
equation for the mean occupation number
\begin{equation}
\langle \dot{\hat n}\rangle= -\Gamma_0[G-D]\langle \hat n\rangle+ \frac{\Gamma_0}{2}\left[\frac{G^2}{4\epsilon}-G+1+ D\right]\,.
\end{equation}
In this expression, the parameter, $D$, describes the heating induced by the recoil kicks from the emitted photons. It is given by
\begin{equation}
D=\eta_g^2 \tilde \alpha + \frac{\Gamma_r}{\Gamma}\left[\frac{\Gamma_g}{\Gamma_g-\Gamma_r}(\eta_r^2\tilde \alpha +\eta_r\eta_g)+
\frac{\Gamma_r}{\Gamma_g-\Gamma_r}\eta_r^2 \right]\,,
\end{equation}
where $ \tilde \alpha=\frac{1}{2}\int N(u) (u-1)^2 du=7/10$. By choosing an appropriate atomic level configuration, $\Gamma_g\gg \Gamma_r$ and/or
$\eta_g^2\gg\eta^2_r$, its value is only limited by $D\approx\eta_g^2\tilde \alpha$. In this case, and for optimized gain the minimal steady state energy is
\begin{equation}
E_{\min}=\hbar \nu \left(\frac{\eta_g^2\tilde \alpha+\sqrt{4\epsilon+\eta_g^4\tilde \alpha^2}}{4\epsilon}\right)\,.
\end{equation}
This expression shows that the minimal energy changes from the Lamb-Dicke to the non-Lamb-Dicke regime at the parameter values, $4\epsilon \approx
\eta_g^2\tilde\alpha$. In the non-Lamb-Dicke regime, i.e., for weak trapping potential the minimal energy approaches the value,
\begin{equation}
E_{\min} \simeq \frac{\tilde \alpha}{\epsilon} \,E_R\,.
\end{equation}
Therefore, the limit for feedback cooling is set by the recoil energy, $E_R$, divided by $\epsilon$, and temperatures well below the Doppler limit, $k_BT_D=\hbar
\Gamma/2$, can be reached.

\section{Feedback vs. EIT Laser Cooling}\label{sec:EITCooling}
In the previous section we focused on the laser detunings $\Delta_p\simeq\Delta_L\simeq 0$ with the goal to measure the atomic momentum to achieve quantum
feedback cooling. As already discussed in Section~\ref{subsec:QFCLD}, in a $\Lambda$-systems EIT laser cooling provides an effective mechanism to cool atoms
essentially to the ground state without any further external manipulation. In this section we derive a master equation which describes both effects, feedback
cooling and ELC, and discuss the cross-over from pure feedback cooling to ELC.

For the adiabatic elimination of the excited states in
Section~\ref{subsec:AE} and therefore for the validity of the
feedback master equation~\eqref{eq:FeedbackME} we required,
$\nu\ll\Omega_{\pm}$. This assumption excludes the parameter regime,
where ELC achieves the lowest temperatures, $\nu\sim
\Omega_+$~\cite{EITCooling_PRA}. In this Section we restrict the
discussion to the Lamb-Dicke limit, $\eta_g,\eta_r\ll 1$. This
allows us to derive a master equation for the motional state for
arbitrary choice of the parameters $\Delta$, $\Omega_L$ and $\nu$.

We start with the full model for the three level atom coupled to
the radiation field as introduced in Section~\ref{sec:Model}. To
optimize the feedback cooling effect and to simplify the following
discussion we make the assumptions, $\Gamma_g=\Gamma$, and as in
the previous sections, $g\ll\Omega_L$. Under the two photon
resonance condition, $\Delta_p=\Delta_L\equiv\Delta$, the system
Hamiltonian, $H_S$, given in Eq.~\eqref{eq:Hamiltonian} can be
written as
\begin{equation}\label{eq:LD_Hamiltonian}
H_S=H_E+H_{I}+H_{\eta}\,.
\end{equation}
The Hamiltonian $H_\eta$ describes the coupling between external
and internal degrees of freedom. Up to first order in the
Lamb-Dicke parameters it is given by
\begin{equation}
H_{\eta}\simeq i\hbar\hat z
\left[\eta_r\frac{\Omega_L}{2}(\sigma_{er}-\sigma_{re})+\eta_g
\frac{g}{2} (\sigma_{eg}-\sigma_{ge})\right]\,.
\end{equation}

The conditioned dynamics of the full atomic density operator, $\rho_c(t)$, is determined by the stochastic master equation~\eqref{eq:ExactME}. As in
Section~\ref{subsec:AE} the goal is to eliminate the internal states and to derive an effective equation for conditioned motional density operator, $\mu_c(t)$.
The principal strategy is the same: For vanishing Lamb-Dicke parameters the decay of the bright states relaxes the atom into the state,
$\rho_c(t)=\mu_c(t)\otimes \pr{D}$. The dynamics of $\mu_c(t)$ can be derived by including the coupling Hamiltonian $H_{\eta}$ in second order perturbation
theory. In contrast to Section~\ref{subsec:AE} we impose no restrictions on the energies of the bright states, $\Omega_\pm$, which leads to resonant transitions
for $|\Omega_\pm|\approx \nu$. Therefore, to guarantee the validity of the perturbation theory we require that $\langle H_\eta\rangle \approx \bar \eta g$ is
much smaller than the decay rates of the bright states, $\Gamma_+\sim \Gamma \cos^2(\theta)$ and $\Gamma_-\sim\Gamma \sin^2(\theta)$.

In Appendix~\ref{app:AE_LambDicke} we use the stochastic Schr\"odinger equation formalism for the adiabatic elimination of the internal states. As a result we
obtain the conditioned master equation,
\begin{equation}\label{eq:SMELambDicke}
\begin{split}
d\mu_c=&-i(\nu+\delta)[ \hat a^\dag \hat a,\mu_c]dt\\
&+ A_-\mathcal{D}[\hat a]\mu_c\,dt+A_+\mathcal{D}[a^\dag ]\mu_c\,dt\\
&+\sqrt{\epsilon\Gamma_0}\mathcal{H}[\hat C
e^{-i\phi}]\mu_c\,dW(t)\,.
\end{split}
\end{equation}
and the expression for the homodyne current,
\begin{equation}\label{eq:CurrentLambDicke}
I_c(t)=\epsilon \Gamma_0\langle \hat C e^{-i\phi}+\hat C^\dag
e^{i\phi}\rangle_c(t)+\sqrt{\epsilon \Gamma_0}\xi(t)\,.
\end{equation}
Here we set $\delta=\Gamma_0{\rm Im}[I(-\nu)+I(\nu)]/4$ and defined the atomic ``jump operator'',
\begin{equation}
\hat C=\frac{\sqrt{2}\nu\Gamma}{\Omega^2}\left[I(-\nu)\hat a+  I(+\nu) \hat a^\dag\right]\,,
\end{equation}
This operator as well as the laser heating and cooling rates, $A_\pm=\Gamma_0{\rm Re}[I(\pm\nu)]/2$, depend on the function $I(\pm\nu)$, which is defined in
Section~\ref{sec:Overview}, Eq.~\eqref{eq:I}.

As in the previous section we consider a feedback force which is
proportional to the measured signal, $I_c(t)$. Note that depending
on values of $I(\pm\nu)$, and the local oscillator phase, $\phi$,
the force is proportional to a linear combination of $\langle\hat
p\rangle$ and $\langle \hat z\rangle$. For the derivation of the
feedback master equation we follow the outline given in
Section~\ref{sec:Feedback} and obtain
\begin{equation}
\begin{split}
\dot \mu=&-i(\nu+\delta) [\hat a^\dag \hat a,\mu]+ A_-\mathcal{D}[\hat a]\mu+A_+\mathcal{D}[a^\dag ]\mu\\
&-i\Gamma_0\frac{G}{2}[\hat z, \hat C e^{-i\phi}\mu+\mu \hat
C^\dag e^{i\phi}]-\Gamma_0\frac{G^2}{8\epsilon}[\hat z,[\hat z
,\mu]]\,.
\end{split}
\end{equation}
Under the rotating wave approximation, which is valid for
$\Gamma_0\ll\nu$, and by neglecting small shifts of the trap
frequency, we end up with master equation
~\eqref{eq:FeedbackME_LD} given in Section~\ref{sec:Overview}.

\begin{figure}
\begin{center}
\includegraphics[width=0.35\textwidth]{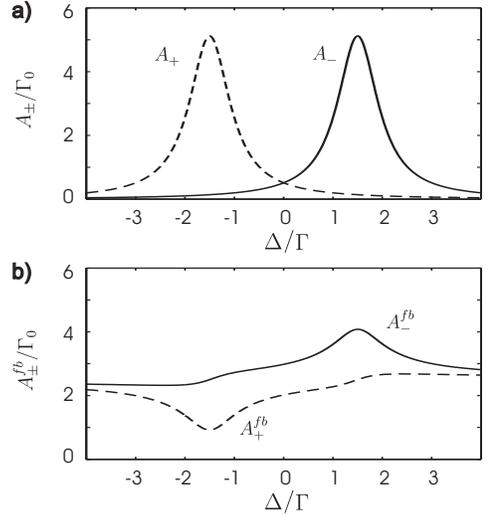}
\caption{ a) Laser heating and cooling rates, $A_\pm$, as a function of the laser detuning, $\Delta$. b) Feedback heating and cooling rates, $A_\pm^{fb}$, for
$G=1$ and an optimized local oscillator phase, $\phi$. The parameters for the two plots are: $\epsilon=0.05$, $\Omega/ \Gamma=0.8$, and $\nu/\Gamma=0.1$. }
\label{fig:LambDickeRates}
\end{center}
\end{figure}

\emph{Discussion:} Fig.~\ref{fig:LambDickeRates} shows the dependence of the four different rates $A_\pm$, $A_\pm^{fb}$ as a function of the detuning $\Delta$.
The cooling and heating rates which originate from the laser interaction, $A_\pm$, correspond to the rates  for ELC derived in Ref.~\cite{EITCooling_PRA}. For
the parameter regime $\Omega> 2\nu$ and for blue detuning, $\Delta >0$, they lead to a minimal temperature for $\Omega^2\approx 4\nu(\nu-\Delta)$. For red
detuning, $\Delta\leq 0$, the heating rate is larger than the cooling rate and without feedback the system does not reach a steady state. By adjusting the phase
$\phi$ the feedback loop always provides additional damping, $W=A_-^{fb}-A_+^{fb}>0$, which for $\Delta=0$ and $\nu\ll \Omega$ is given by $W=G\Gamma_0$. The
noise added by the feedback loop $\Gamma_0 G^2/8\epsilon$, imposes a restriction on $G$ if one is interested in low steady state energies. The combined effect of
feedback and EIT laser cooling lead to a final temperature which is plotted in Fig.~\ref{fig:GainPlot} in Section~\ref{subsec:QFCLD}.

\section{Conclusion}
In this paper we have shown that a continuous readout of the momentum of a single atom can be achieved by employing the high velocity sensitivity of the index of
refraction in a driven $\Lambda$-system. The transparency effect for an atom at rest and the linear dependence of the index of refraction on the Doppler shift
lead to a homodyne current linear in $\langle \hat p\rangle$ and a minimal back action on the atomic motion, approaching the quantum limit for
$\epsilon\rightarrow 1$.

By applying a force which is proportional to the measured signal, feedback cooling for single ions can be realized.  The cooling scheme is applicable in and
outside the Lamb-Dicke regime with steady state temperatures well below the Doppler limit. From a fundamental point of view we want to point out, that in the
proposed feedback scheme the measured current is fed back \emph{directly} on the trap electrodes. Therefore, its implementation allows for a test of the theory
of direct quantum feedback~\cite{Feedback_PRL} on an individual quantum system close to the ground state.

\begin{acknowledgments}
The authors thank  G. Morigi, J. Eschner and the experimental group around R. Blatt for useful discussions. Work at Innsbruck was supported in part by the
Austrian Science Foundation FWF, European Networks and the Institute for Quantum Information.
\end{acknowledgments}




\begin{appendix}
\section{Adiabatic elimination}\label{app:AE_General}
In this appendix we derive the stochastic master equation~\eqref{eq:PrelimSME} for the external density operator, $\mu_c$, and the expression for the signal
$I_c(t)$~\eqref{eq:PrelimCurrent}. We start with the conditioned master equation given in Eq.~\eqref{eq:ExactME} and apply the unitary transformation as defined
in Eq.~\eqref{eq:UnitaryTransformation}. In the new basis the stochastic master equation can be written in the form
\begin{equation}\label{eq:TransFullSME}
\begin{split}
d\tilde \rho_c=\left(\mathcal{L}_I+\mathcal{L}_\nu+
\tilde{\mathcal{J}}\right)\tilde \rho_c dt
+\sqrt{\epsilon\Gamma_g}\mathcal{H}[\sigma_{ge}e^{-i\phi}]\tilde
\rho_c dW(t) \,,
\end{split}
\end{equation}
where we divided the total Liouville operator into the three
contributions,
\begin{eqnarray*}
\mL_{I}(\rho)&=&-\frac{i}{\hbar}[H_I,\rho]-\frac{\Gamma}{2}(\pr{e}\rho+\rho \pr{e})\,,\\
\mL_{\nu}(\rho)&=&-\frac{i}{\hbar}[H_E+H_{\rm int },\rho]\,,\\
\tilde{\mathcal{J}}(\rho)&=&\Gamma\tilde{\mathcal{J}_g}(\sigma_{ge}\rho \sigma_{eg})+\Gamma\tilde{\mathcal{J}_{r}}(\sigma_{re}\rho\sigma_{er})\,,.\\
\end{eqnarray*}
The action of the recycling operators,
$\tilde{\mathcal{J}}_{g,r}$, is defined in Section~\ref{sec:Model}
Eq.~\eqref{eq:RecyclingTerms}.

In the following we write the total density operator in terms of
the eigenbasis of $H_I$ as, $\tilde \rho_c=\sum_{i,j} \tilde
\mu_{ij}\otimes|i\rangle\langle j|$ with $i,j\in\{+,-,D\}$.
Reinserting this decomposition into Eq.~\eqref{eq:TransFullSME} we
obtain a set of coupled equations for the external operators
$\tilde \mu_{ij}$. By grouping the 9 elements $\tilde \mu_{ij}$
into a single vector $\vec \mu$, the resulting set of equations
can be written in the form
\begin{equation}\label{eq:dvecmu}
d \vec \mu=(\mathbf{L}_I+\mathbf{L}_\nu+\mathbf{J})\vec \mu\, dt+
 \sqrt{\epsilon \Gamma_g}\left(\mathbf{S} -{\rm Tr}\{\vec V\! \cdot\! \vec\mu\}\right)\vec
\mu \,dW(t)\,.
\end{equation}
The entries of the matrices $\mathbf{L}_I$, $\mathbf{L}_\nu$,
$\mathbf{J}$, $\mathbf{S}$ and the vector, $\vec V$, can be
derived in a straight forward (but lengthy) way by writing the
operators in Eq.~\eqref{eq:TransFullSME} in terms of the states,
$|+\rangle$, $|-\rangle$ and $|D\rangle$. Note that the entries of
$\mathbf{L}_\nu$ and $\mathbf{J}$ still contain superoperators
acting on the external operators $\tilde \mu_{ij}$.

The goal is to derive an effective equation for the population in the dark state, $\tilde \mu_0:=\tilde \mu_{DD}$, for the parameter regime $\nu,\Delta_D\ll
\Omega_\pm$. Formally we can combine both conditions and make a series expansion in the trap frequency, $\nu$.  According to the structure of $\mathbf{L}_I$ we
group the external operators $\tilde \mu_{ij}$ into the two vectors $\vec \mu_1:=(\tilde \mu_{D+},\tilde\mu_{+D},\tilde\mu_{D-},\tilde\mu_{-D})^{T}$ and $\vec
\mu_2:=(\tilde\mu_{++},\tilde\mu_{--},\tilde\mu_{+-},\tilde\mu_{-+})^{T}$. By ordering the entries of $\vec \mu$ such that, $\vec \mu =(\vec \mu_2^T,\vec
\mu_1^T,\mu_0)^T$, the matrices $\mathbf{L}_I$, $\mathbf{J}$ and $\mathbf{L}_\nu$ have the block form
\begin{equation*}
\mathbf{L}_I+\mathbf{J} = \left(\begin{array}{ccc} L_I^{2}+J^{2} & 0 & 0  \\ J^{12} & L_I^{1} & 0 \\
J^{02} & 0 & 0\end{array}\right),
\mathbf{L}_\nu = \left(\begin{array}{ccc} L_\nu^{2} & L^{21}_\nu & 0  \\ L^{12}_\nu & L_\nu^{1} & L_{\nu}^{10} \\
0 & L_\nu^{01} & L_\nu^{0}\end{array}\right).
\end{equation*}
From the structure of $\mathbf{L}_I$ and $\mathbf{L}_\nu$ we see
that for $\nu\rightarrow 0$ we have $\vec
\mu_1\sim\mathcal{O}(\nu)$ and $\vec \mu_2\sim\mathcal{O}(\nu^2)$.
Therefore, up to second order in $\nu$ the equation for $\tilde
\mu_0$ is
\begin{equation}\label{eq:dRho0}
\begin{split}
d\tilde \mu_0=&\left(L_\nu^{0}\tilde \mu_0+L_\nu^{01}\vec \mu_1+J^{02}\vec \mu_2\right)dt\\
  &+\sqrt{\epsilon \Gamma_g}\left(\vec v \cdot\vec\mu_1-{\rm Tr}\{\vec v\cdot\vec \mu_1\}\mu_0\right)dW(t)\,,
\end{split}
\end{equation}
with
\begin{equation}
\vec v =\frac{\Omega_L}{\Omega}(e^{i\phi}
\cos(\theta),e^{-i\phi}\cos(\theta),e^{i\phi}\sin(\theta),e^{-i\phi}\sin(\theta))\,.
\end{equation}
 The equations for $\vec \mu_1$ and $\vec \mu_2$ are given by
\begin{eqnarray*}
\dot{\vec \mu}_1&=&L_I^1\vec \mu_1 +L_{\nu}^{10}  \tilde \mu_0+\mathcal{O}(\nu^2)\,,\\
\dot{\vec \mu}_2&=&\left(L_{I}^2+ J^{2}\right)\vec \mu_2
+L_{\nu}^{21} \vec \mu_1 +\mathcal{O}(\nu^3)\,.
\end{eqnarray*}
They can be integrated and up to the relevant orders of $\nu$ we obtain the formal solution
\begin{eqnarray*}
\vec \mu_1&=&(L_I^1)^{-1}L_{\nu}^{10}  \tilde \mu_0+\mathcal{O}(\nu^2)\,,\\
\vec \mu_2&=&\left(L_{I}^2 + J^{2}\right)^{-1}L_{\nu}^{21}
(L_{I}^1)^{-1}L_{\nu}^{10} \tilde \mu_0 +\mathcal{O}(\nu^3)\,.
\end{eqnarray*}
Resubstituting these expressions into Eq.~\eqref{eq:dRho0} the
resulting equation can be written in the form
\begin{equation}\label{eq:drho0}
\begin{split}
d\tilde \mu_0=&-i[ \hat h_{\rm eff}\tilde \mu_0-\tilde \mu_0  \hat h_{\rm eff}^\dag]dt  +  \lambda^2 \Gamma \mathcal{R} (\hat p \tilde \mu_0 \hat p) dt \\
&+\sqrt{\frac{\epsilon\Gamma_g\lambda^2\Omega_L^2}{\Omega^2}}\mathcal{H}[\hat p e^{-i\phi}]\tilde \mu_0 dW(t)\,,
\end{split}
\end{equation}
with $\lambda = 2 \bar \eta \nu g \Omega_L/\Omega^3$, the
non-hermitian operator
\begin{equation}
\hat h_{\rm eff}=\nu a^\dag a-\frac{g\lambda}{2}\hat p-\lambda^2 \Delta \hat p^2-i\frac{\lambda^2\Gamma}{2}\hat p^2\,.
\end{equation}
and the ``recycling'' term,
\begin{equation}
\mathcal{R}=
\frac{\frac{\Omega_L^2}{\Omega^2}\tilde{\mathcal{J}}_g +
\frac{g^2}{\Omega^2}\tilde{\mathcal{J}}_r}{1- \frac{g^2}{\Omega^2}
 \tilde{\mathcal{J}}_g-\frac{\Omega_L^2}{\Omega^2}\tilde{\mathcal{J}}_r}\,.
\end{equation}
In the formal expression for $\mathcal{R}$ the inversion of
operator is justified in the limit of a weak probe field,
$g\ll\Omega_L$ and $\Gamma_r\ll\Gamma$. From the stochastic term
in Eq.~\eqref{eq:drho0} we see that these conditions also maximize
the signal strength for a given decoherence rate,
$\lambda^2\Gamma$.

In the original basis the evolution for $\mu_c$ is given by the
relation,
\begin{equation}
d\mu_c={\rm Tr}_I\{U d\tilde \mu_0\otimes \pr{D}U^\dag\}.
\end{equation}
Due to the overlap $|\langle r |D\rangle|^2=g^2/ \Omega^2$ the action of $U$ on the external states reduces to the action of the operator $\exp(i\bar \eta \hat z
g^2/\Omega^2)$. Therefore, the only effect of the basis transformation is the cancellation of the term $\lambda g \hat p/2$ in the effective Hamiltonian, $\hat
h_{\rm eff}$.

For the expression of the measured signal $I_c(t)$~\eqref{eq:PrelimCurrent} we can simply repeat the calculations from above.  Using the same notation as in
Eq.~\eqref{eq:dvecmu} it can be written as
\begin{equation}
I_c(t)= \epsilon \Gamma_g {\rm Tr}\{\vec V \!\cdot\! \vec \mu\}
+\sqrt{\epsilon \Gamma_g}\xi(t)\,.
\end{equation}
We see that the first term already appeared in the stochastic
master equation~\eqref{eq:dvecmu} and can be evaluated along the
same lines. By multiplying the resulting expression by a factor
$\lambda$, we obtain the current given in
Eq.~\eqref{eq:PrelimCurrent}.

\section{Adiabatic elimination in the Lamb-Dicke limit}\label{app:AE_LambDicke}
In this appendix we derive the conditioned evolution of the
external atomic state in the Lamb-Dicke limit, $\eta_p,\eta_L\ll
1$. We start with the stochastic Schr\"odinger equation for the
total wavefunction, $|\Psi\rangle$, which includes the state of
the atom as well as the state of the electromagnetic environment.
For the system Hamiltonian $H_S$~\eqref{eq:LD_Hamiltonian} and the
atom-field interaction described in Section~\ref{sec:Model} it is
given by
\begin{equation}
\begin{split}
d|\Psi\rangle=&\left(-\frac{i}{\hbar} H_S - \frac{\Gamma}{2}\pr{e}\right)|\Psi\rangle dt\\
 +&\sqrt{\Gamma}\int_{-1}^1 du \sqrt{N(u)}\,e^{-i\eta_p\hat z u}\sigma_{ge}dB_u^\dag(t)|\Psi\rangle\,.
\end{split}
\end{equation}
The noise increment operators, $dB_u^\dag(t)$, fulfill the Ito
rules $dB_u(t) dB_{u'}^\dag(t)=\delta(u-u')dt$ and correspond to
the emission of  photons under an angle $\alpha=\arccos(u)$ with
respect to the z-axis.

 We decompose the total wave function in terms of the eigenstates of $H_I$ as $|\Psi\rangle=\sum_{i} |\Psi_i\rangle \otimes |i\rangle$, with $i=+,-,D$.
 For vanishing Lamb-Dicke parameters, $\eta_j\rightarrow 0$, and after some transient deviations the system evolves into the state,
 $|\Psi\rangle=|\Psi_D\rangle\otimes|D\rangle$.
  The coupling between the external and internal degrees of
freedom, $H_{\rm int}$, leads to finite contributions from the
bright states which are of the order of the Lamb-Dicke parameter,
$|\Psi_\pm\rangle\sim \mathcal{O}(\bar \eta)$. In the following we
treat the coupling to the excited states in perturbation theory to
derive an effective equation for $|\Psi_D\rangle$ which is valid
up to second order in $\bar \eta$.

In the interaction picture with respect to the external
Hamiltonian, $H_E=\hbar \nu \hat a^\dag \hat a$, the equation for
the dark state wave function is
\begin{equation}\label{eq:DarkStateEquation}
d|\Psi_D\rangle\simeq-\frac{\bar \eta g}{2} \hat z(t)
|\Psi_e\rangle dt +\sqrt{\Gamma}\int_{-1}^1 du
\sqrt{N(u)}dB_u^\dag(t)|\Psi_e\rangle\,,
\end{equation}
with $|\Psi_e\rangle=\cos(\theta)|\Psi_+\rangle
+\sin(\theta)|\Psi_-\rangle$. Note that in this equation we
already used the assumption of a weak probe field, and set
$\Omega=\sqrt{\Omega_L^2+g^2}\approx \Omega_L$. Under the same
assumption the equations for the bright states are given by
\begin{equation}\label{eq:ExcitedStates}
\frac{d}{dt}\left(\begin{array}{c} |\Psi_+\rangle \\|\Psi_-\rangle
\end{array}\right) = -\mathbf{M}\left(\begin{array}{c}
|\Psi_+\rangle \\|\Psi_-\rangle
\end{array}\right) + \frac{\bar \eta g}{2} \hat z(t)\left(\begin{array}{c} \cos(\theta) \\\sin(\theta) \end{array}\right)|\Psi_D\rangle\,,
\end{equation}
where we defined the matrix
\begin{equation}
\mathbf{M}=\left(\begin{array}{cc} i\Omega_+
+\frac{\Gamma}{2}\cos^2(\theta) & \frac{\Gamma}{4}\sin(2\theta)
\\ \frac{\Gamma}{4}\sin(2\theta)& i\Omega_-
+\frac{\Gamma}{2}\sin^2(\theta)
\end{array}\right)\,.
\end{equation}
Up to first order in $\bar \eta$, the solution for the excited
states is
\begin{equation}
\left(\begin{array}{c} |\Psi_+\rangle \\|\Psi_-\rangle
\end{array}\right) = \frac{\bar \eta  g}{2} \left[\int_{-\infty}^t
e^{-\mathbf{M}(t-s)} \hat z(s)ds\right]\left(\begin{array}{c}
\cos(\theta) \\\sin(\theta) \end{array}\right)|\Psi_D\rangle.
\end{equation}
By inserting the time dependence of the position operator, $\hat
z(t)=(\hat a e^{-i\nu t}+\hat a^\dag e^{i\nu t})/\sqrt{2}$, we can
evaluate this integral and obtain the evolution of the
wavefunctions, $|\Psi_\pm\rangle$.  This solution is then
reinserted into Eq.~\eqref{eq:DarkStateEquation} to get the
equation of the dark state wave function up to order $\bar
\eta^2$. As long as the final dynamics in the interaction picture
is slow compared to the trap frequency, $\nu$, we can use the
rotating wave approximation and neglect terms proportional to
$e^{\pm i 2 \nu t}$. The resulting equation is then given by
\begin{equation}\label{eq:SSE_DarkState}
\begin{split}
d|\Psi_D\rangle = -&\frac{\bar \eta^2 g^2}{8}\left[ \tilde I(\nu)\hat a\hat a^\dag+ \tilde I(-\nu) \hat a^\dag \hat a\right]|\Psi_D\rangle dt\\
                +& \sqrt{\frac{\bar \eta^2 g^2}{8}\Gamma}\int_{-1}^1 du \sqrt{N(u)}\,\tilde C(t) dB^\dag_u(t) |\Psi_D\rangle\,,
\end{split}
\end{equation}
where we set $\tilde  C(t)=(\tilde I(-\nu)\hat a(t)+\tilde I(\nu)\hat a^\dag(t))$, and defined the function $\tilde I(\nu)$ by
\begin{equation}
\tilde I(\nu):=(\cos(\theta),\sin(\theta))\left[\int_{0}^\infty
e^{-\mathbf{M}\tau} e^{-i\nu\tau}
d\tau\right]\left(\begin{array}{c} \cos(\theta) \\ \sin(\theta)
\end{array}\right)\,.
\end{equation}

Apart from the motional state of the atom, the wavefunction $|\Psi_D\rangle$ still incudes the full state of the electromagnetic environment. To obtain the
conditioned dynamics for the external density operator, $\mu_c$, we first decompose the set of noise increment operators into the two contributions,
\begin{eqnarray*}
 dB_p^\dag(t)&=&\frac{1}{\sqrt{\epsilon}}\int_{1-\epsilon}^{1} du \sqrt{N(u)}\,dB_u^\dag(t)\,,\\
 dB_{\perp}^\dag(t) &=& \frac{1}{\sqrt{1-\epsilon}} \int_{-1}^{1-\epsilon} du \sqrt{N(u)}\,dB_u^\dag(t)\,  \,.
\end{eqnarray*}
As in Section~\ref{sec:Model} the parameter, $\epsilon$,
determines the fraction of the photons which are scattered into
the mode of the probe beam. The increment operator $dB_p^\dag(t)$
obeys the Ito rule, $dB_p(t)dB_p^\dag(t)=dt$, and corresponds to
the emission of photons, which are focused on the detector.

A rigorous, but rather technical way to convert the stochastic
Schr\"odinger equation into a stochastic master equation can be
found in Ref.~\cite{QuantumNoise, Graham}. A convenient shortcut
is, to first derive the unconditioned master equation for
$\mu={\rm Tr}_{EM}\{|\Psi_D\rangle\langle \Psi_D|\}$ and then
perform a partial 'unravelling' of this master equation as
discussed in Ref.~\cite{Wiseman_Homodyne}. For a phase $\phi$ of
the local oscillator  we finally get the equation
\begin{equation}
\begin{split}
d\mu_c= &-i \frac{\bar \eta^2 g^2}{8}{\rm Im}[\tilde
I(-\nu)+\tilde I (+\nu)]\,[\hat a^\dag \hat a,\mu_c] dt
\\&+\frac{\bar \eta^2 g^2}{4}\left({\rm Re}[\tilde I(-\nu)]\mathcal{D}[\hat a]
         +{\rm Re}[\tilde I(+\nu)]\mathcal{D}[\hat a^\dag]\right) \mu_c dt\\
         &+\sqrt{\epsilon \Gamma \frac{\bar \eta^2 g^2}{8}}\,\mathcal{H}[\tilde C(t) e^{-i\phi}]\mu_c\, dW(t)\,,
\end{split}
\end{equation}
 The measured current has the form
\begin{equation}
I_c(t)=\epsilon \Gamma \frac{\bar \eta^2 g^2}{8}\langle \tilde C(t) e^{-i\phi} +\tilde C^\dag(t) e^{i\phi}\rangle_c +\sqrt{\epsilon \Gamma \frac{\bar \eta^2
g^2}{8}}\xi(t)\,.
\end{equation}
To make a comparison to results of Section~\ref{sec:Model} and
Section~\ref{sec:Feedback} we introduce the rescaled function,
$I(\nu)$, by setting $ I(\nu)=\tilde
I(\nu)\Omega^4/(8\nu^2\Gamma)$. With this definition and a
rescaling of the current we finally obtain
Eqs.~\eqref{eq:SMELambDicke} and~\eqref{eq:CurrentLambDicke}.

\end{appendix}




\bibliographystyle{apsrev}
\bibliography{ref}

\end{document}